\documentstyle[epsfig] {article}

\begin{document}

\flushbottom
 
\begin{titlepage}
 
\begin{tabbing}
\` UT-ICEPP 03-06 \\
\` July 2003 \\
\end{tabbing}

\bigskip
\bigskip
 
\begin{center}{\LARGE\bf 
Solution of Orthopositronium lifetime Puzzle
}\end{center}

\bigskip

\begin{center}{ 
Contribution to Proceedings of \\
Workshop on Positronium Physics\\
ETH Honggerberg Zurich (May 30-31 2003)\\
}\end{center}

\bigskip

\begin{center}{\large 
S. Asai~\footnote[1]{Mailing address: CERN EP-Div, CH-1211, Geneva 23, 
Switzerland, E-mail address: Shoji.Asai@cern.ch}, 
O. Jinnouchi and T. Kobayashi
}\end{center}

\bigskip

\bigskip

\begin{abstract}
{
The intrinsic decay rate of orthopositronium formed in ${\rm SiO_2}$ 
powder is measured using the direct 
$2\gamma$ correction method such that the time dependence of the 
pick-off annihilation rate is precisely determined.
The decay rate of orthopositronium is 
found to be $7.0396\pm0.0012 (stat.)\pm0.0011 (sys.)\mu s^{-1}$, 
which is consistent with our previous measurements with about twice the accuracy. 
Results agree well with the $O(\alpha^2)$ QED prediction, and also with a result
reported very recently using nanoporous film. 
}
\end{abstract}

\bigskip
\bigskip

\begin{center}{\large ICEPP}\end{center}
\begin{center}
{\large International Center for Elementary Particle Physics,}
\end{center}
\begin{center}{\large University of Tokyo}\end{center}
\begin{center}{\large 7-3-1 Hongo, Bunkyo-ku, Tokyo 113-0033, Japan}\end{center}

\bigskip

\end{titlepage}

\section{History: Orthopositronium lifetime puzzle}\label{sec:intro}

Positronium (Ps), the bound state of 
an electron and a positron,
is a purely leptonic system,
and the triplet ($1^{3}S_{1}$) state of Ps, orthopositronium(o-Ps),
decays slowly into three photons. 
Precise measurement of this decay rate gives
us direct information about quantum electrodynamics(QED) in
bound state.

Three precision measurements\cite{GAS87,GAS89,CAV90} 
of the o-PS decay rate were performed at Ann Arbor, 
which reported decay rate values much larger, i.e., 5.2 -- 9.1 experimental 
standard deviations, than a QED prediction\cite{ADKINS-4}
($7.039934(10)~\mu s^{-1}$) corrected up to $O(\alpha^2)$\@.
This discrepancy has been referred as `orthopositronium lifetime puzzle', 
and was long-standing problem.
To elucidate discrepancies, a variety of experiments have since 
been carried out to search for the exotic 
decay mode of o-Ps, resulting in no evidence so far 
\cite{EXOTIC-LL,EXOTIC-SL,EXOTIC-IV,EXOTIC-UB,EXOTIC-TW,EXOTIC-FOUR}\@.

As some fraction of o-Ps inevitably results in `pick-off' annihilations 
due to collisions with atomic electrons of the target material, 
the observed o-Ps decay rate 
$\lambda_{obs}$ is a sum of the intrinsic o-Ps 
decay rate $\lambda_{{\rm o}\mbox{-}{\rm Ps}}$ and the pick-off 
annihilation rate into $2\gamma$'s, 
$\lambda_{pick}$, i.e.,

\begin{equation}
\lambda_{obs}(t)=\lambda_{3\gamma}+\lambda_{pick}(t).
\end{equation} 

$\lambda_{pick}(t)$ is proportional 
to the rate of o-Ps collisions with the target materials, i.e.; 
$\lambda_{pick}=n\sigma_a v(t)$, where $n$ is 
product of the density of the target, $\sigma_a$ the annihilation cross-section, 
and $v(t)$ the time dependent velocity of o-Ps. 
Due to the thermalization process of o-Ps, 
this necessitates expressing 
$\lambda_{pick}$ as a function of time whose properties are dependent on 
the surrounding materials. 
Thermalization process should be carefully treated even in the 
cavity experiment\cite{CAV90}.
Although pickoff correction is small in cavities, 
disappearance of o-Ps through the cavity entrance aperture 
has large contribution to $\lambda_{obs}$.
This disappearance rate is also proportional to $v(t)$, as the same reason.
Since the rate of elastic collision is extremely small in cavities, 
it takes much time, longer than 1 $\mu s$,
to thermalize well, and the disappearance rate still depends 
strongly on time.

In previous measurements\cite{GAS87,GAS89,CAV90}, 
$\lambda_{obs}$'s were measured by varying the 
densities of the target materials, size of the cavities and
also the entrance aperture of the cavities.
The extrapolation to zero density or aperture was expected to 
yield the decay rate in a vacuum, $\lambda_{3\gamma}$, under 
the assumption of quick thermalization (shorter than 170-180 nsec)
with constant o-Ps velocity. 
However, this assumption contains a serious systematic error as pointed out in 
reference\cite{PEKIN,ASAI95}\@.

We have proposed the following entirely new method\cite{ASAI95}, 
which is free from above-mentioned systematic error.
The energy distribution of photons from the 3-body decay is 
continuous below the steep edge at 511~keV, 
whereas the pick-off annihilation is 2-body which produces 
a 511~keV monochromatic peak. 
Energy and timing information are simultaneously
measured with high-energy resolution germanium detectors such 
that $\lambda_{pick}(t)/\lambda_{3\gamma}$ can be determined
from the energy spectrum of the emitted photon. 
Once a precise thermalization function is obtained, $\lambda_{pick}(t)$ will 
contain all information about the process. 
The population of o-Ps at time $t$, 
$N(t)$ can be expressed as \begin{equation}
N(t)=N_0' \exp\left(-
\lambda_{3\gamma}\int^{t}_0\left(1+\frac{\lambda_{pick}(t')}
{\lambda_{3\gamma}}\right)dt'\right).
\end{equation}
Providing the ratio is determined as a function of time,
the intrinsic decay rate of o-Ps, $\lambda_{3\gamma}$, can be directly obtained
by fitting the observed time spectrum.

We obtained decay rate of $7.0398(29)$
and $7.0399(25)~\mu s^{-1}$ independently\cite{ASAI95,JIN00}, 
which are consistent with the non-relativistic QED calculation\cite{ADKINS-4}, 
and quite differ from the results obtained at Ann Arbor\cite{GAS87,GAS89,CAV90},
7.0482(16)--7.0516(13)$~\mu s^{-1}$\@.
The observed $\lambda_{pick}(t)$ indicates that o-Ps thermalization
is slow and it is serious systematic problem in all experiments using an extrapolation.   
In 1998, they recognized that the incomplete thermalization makes 
the serious problem in their results\cite{Ann1}, but they did not update results, 
and the discrepancy still remaining. 

There are also still several problems remained in our result\cite{ASAI95}: 
(i) accuracy was $350~ppm$, being worse than those of the other
experiments\cite{GAS87,GAS89,CAV90}, (ii) There was unknown systematic uncertainties before
$t_{start}=200~ns$, and therefore to remove this uncertainty, 
final results were obtained using data after $220~ns$,  
and (iii) systematic error regarding the Stark effect was not estimated.
Improving results by considering these problems is described\cite{D-thesis} here,
and we report the final result using our new method.

\section{Experiment}\label{sec:experiment}

\begin{figure}[t!]
\vspace{0.2pc}
  \begin{center}
    \includegraphics[width=14pc, angle=-90,keepaspectratio]{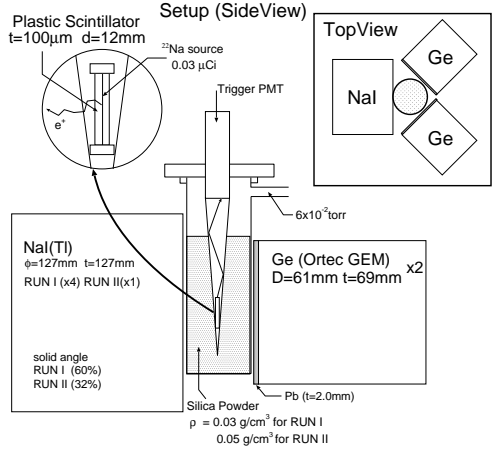}
  \end{center}
\caption{Schematic diagram of experimental setup.}
\label{fig:setup}
\end{figure}
Figure~\ref{fig:setup} shows a diagram of the experimental setup;
A ${}^{22}{\rm Na}$ positron source (dia., $2~mm$) with approximate 
strength of $0.03\mu Ci$, being sandwiched between two sheets of plastic 
scintillators and held by a cone made of aluminized mylar. 
The cone was situated at the center of a cylindrical $50~mm$-diameter 
vacuum container made of $1~mm$-thick glass, being filled with ${\rm 
SiO_2}$ powder and evacuated down to $5\times10^{-2}$ Torr. Two different types of 
${\rm SiO_2}$ powder were used as listed in Table~\ref{table:powder}, 
with the biggest difference being in the mean distance between grains 
such that different pick-off ratio would be obtained. Using these 
powders, two 6-month runs were performed.

\renewcommand{\arraystretch}{1.00}
\begin{table}[htb]
\begin{center}
\begin{tabular}{crr}\hline
                   & RUN1 & RUN2 \\ \hline\hline
primary grain size (nm)&        7  &       7 \\ 
surface area  ($m^2/g$)&   $300\pm30$ & $260\pm30$ \\
density ($g/cm^3)$       &   0.03 & 0.05 \\ 
mean distance between grains (nm)     &   340  & 200 \\ 
surface & hydrophile & hydrophobe \\ \hline
\end{tabular}
\caption{Characteristics of ${\rm SiO_2}$ powders used in the measurements.}
\label{table:powder}
\end{center}
\end{table}
Most of the emitted positrons pass through the scintillator,
transmitting a light pulse to a 
photomultiplier(Trigger PMT) 
and forming Ps when stopped in the silica powder.
Two high-purity coaxial germanium detectors (Ortec GEM38195) 
precisely measured the thermalization process.
Energy resolutions were measured using several line $\gamma$ sources, 
with typical resultant values of $0.53$ and $0.64~{\rm keV}$ at 514~keV\@. 
Lead sheets 2.0-$mm$-thick were placed in front of each 
detector for suppressing contributions from simultaneous low-energy $\gamma$ hits from the 
3$\gamma$ decay of o-Ps.
Four large cylindrical NaI(Tl) scintillators 
(Scionix 127A127/5; 127~$mm~(\phi)$ $\times$ 127~$mm~(t)$) 
simultaneously measured the time and energy information from each decay. 

A new time-to-digital converter (TDC) was developed and employed 
for the present measurement. 
This direct clock-counting type TDC(2GHz) has a time-resolution of $0.5~nsec$ 
with known accuracy of 1-$ppm$. 
The time range for each channel is $32~\mu s$, and the integral non-linearity,
which is an important source of systematic errors,
is controlled to be extremely small at $<15~ppm$.
To provide a systematic check, we 
used a 200-MHz internal clock-based TDC with 5~$ns$ time resolution.

The trigger PMT signal is fed into a fast leading discriminator 
whose output provides common start signals for the TDCs.
Each output signal from the photo-multipliers to the NaI(Tl)s is fed into 
three ADCs and a fast leading edge discriminator
that provides stop signals for the TDCs and gate timing 
for the ADCs.  
One ADC with a 3-$\mu s$ gate width, measures the 
whole charge for the duration of the signal. 
The other two ADCs eliminate pile-up events at the tail of the signal 
and base-line fluctuations, respectively. 
One ADC measures the signal charge with a narrower gate width (250\ $ns$) 
comparable to the intrinsic decay 
time constant of the NaI(Tl) scintillator, while another
ADC measures the base-line condition of the signal (180\ $ns$ width) 
just prior to the event. 
All gates for these ADCs are individual and gate timings 
are optimized for signal timings.

Signal outputs from the Ge detectors are used for timing measurements
and obtaining precise energy information. 
One signal is fed into a fast-filter amplifier (FFA)
whose output is used as the stop signal for the TDCs, two auxiliary 
ADC signals, and three discriminaters whose thresholds are 
set at $-50, -100$, and $-150$~mV. 
Each FFA output is fed into another high-resolution TDC having 
fine resolution of $250~ps$. 
Signal timings are determined utilizing signal shapes calculated from different threshold 
discriminating times and extrapolating to the intrinsic timing\cite{D-thesis}. 
Good time resolution of $4~ns$ is obtained, as is efficient
rejection power for slow rise signal components known to disturb Ge timing spectroscopy. 
Similar to the NaI(Tl) detectors, the two auxiliary ADCs effectively reject pile-up 
events by measuring the signal charge with a narrower gate
and earlier timing gate. 
Precise time and energy resolutions can be obtained with this rejection scheme. 
The other output from each Ge detector is independently fed
into a spectroscopy amplifier. 
Amplified with a 6-$\mu s$ time constant, output is provided to a Wilkinson type peak-
holding ADC that provides a precise energy spectrum. 

\section{Analysis and discussion}\label{sec:analysis}

The ratio $\lambda_{pick}(t)/\lambda_{3\gamma}$ is 
determined using the energy spectrum measured by 
the Ge detectors. The energy spectrum of the o-Ps decay sample, 
referred to as the {\it o-Ps spectrum}, is 
obtained by subtracting accidental contributions 
from the measured spectrum. 

The $3\gamma$-decay continuum spectrum is calculated using Monte Carlo
simulation in which the geometry and various material distributions
are reproduced in detail.
For every simulated event, three photons are generated according to 
an order-$ \alpha $-corrected energy spectrum\cite{Adkins_Private}.
Successive photoelectric, Compton, or Rayleigh scattering interactions 
of every photon are then followed through the materials 
until all photon energy is either deposited or escapes  from the detectors. 
The response function of the detectors is determined based on the measured spectrum of 
monochromatic $\gamma$-rays emitted from ${}^{152}{\rm Eu}$, ${}^{85}{\rm Sr}$, and 
${}^{137}{\rm Cs}$, with this function being used in the simulation.
We refer to the obtained spectrum as 
the {\it $3\gamma$-spectrum} which is normalized to the o-Ps spectrum with 
the ratio of event numbers  within the region (480-505~keV). 

\begin{figure}[t!]%
\vspace{0.2pc}
  \begin{center}
    \includegraphics[width=15pc,keepaspectratio]{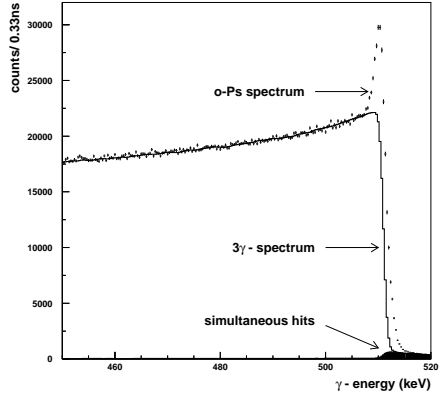}
    \includegraphics[width=15pc,keepaspectratio]{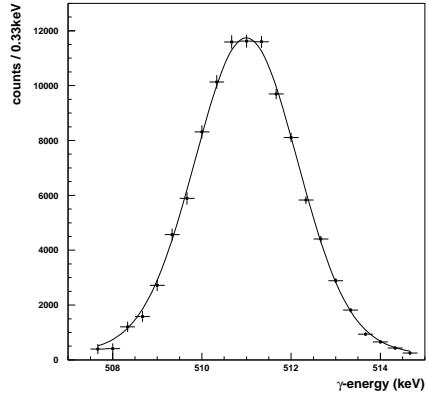}
  \end{center}
\caption{(a) Energy spectrum of o-Ps decay $\gamma$'s obtained by Ge detectors. 
Dots represent data points in a time window of $150-700~ns$, and the solid 
line shows the $3\gamma$-decay spectrum calculated by Monte Carlo 
simulation. Shaded area indicate simultaneous hits estimated by the simulation. 
(b) Pick-off spectrum obtained after subtracting the $3\gamma$ contribution 
from the o-Ps spectrum. The solid line represents the fit result.}
\label{fig:ops-3gamma}\
\end{figure}

Figure \ref{fig:ops-3gamma}(a) shows good agreement between the o-Ps spectrum and 
$3\gamma$ spectrum below 508~keV, where the pick-off annihilation peak is 
evident at the edge of the $3\gamma$-decay continuum. 
The lead sheets in front of the Ge detectors effectively 
suppress the contribution from events in which two 
low-energy $\gamma$'s emitted from a $3\gamma$-
decay simultaneously hit one detector and that the $3\gamma$-spectrum well reproduces such 
simultaneous events.
Figure \ref{fig:ops-3gamma}(b) shows an enlarged view of the observed o-Ps
spectrum after subtracting the $3\gamma$-spectrum, 
where good agreement with a detector response function  
is present.
The centroid of this spectrum is $510.997~^{+0.003}_{-
0.024}~\mbox{(keV)}$, which is consistent with $511.0$~(keV).
This indicates successful subtraction of the $3\gamma$ contribution, and 
the resultant peak can be regarded as pure pick-off annihilation samples.
Obtained ratio of $\lambda_{pick}/\lambda_{3\gamma}$ is $0.01049(8)$ 
for time window of $150-700~ns$\@.

\begin{figure}[t!]%
\vspace{0.2pc}
  \begin{center}
    \includegraphics[width=20pc,keepaspectratio]{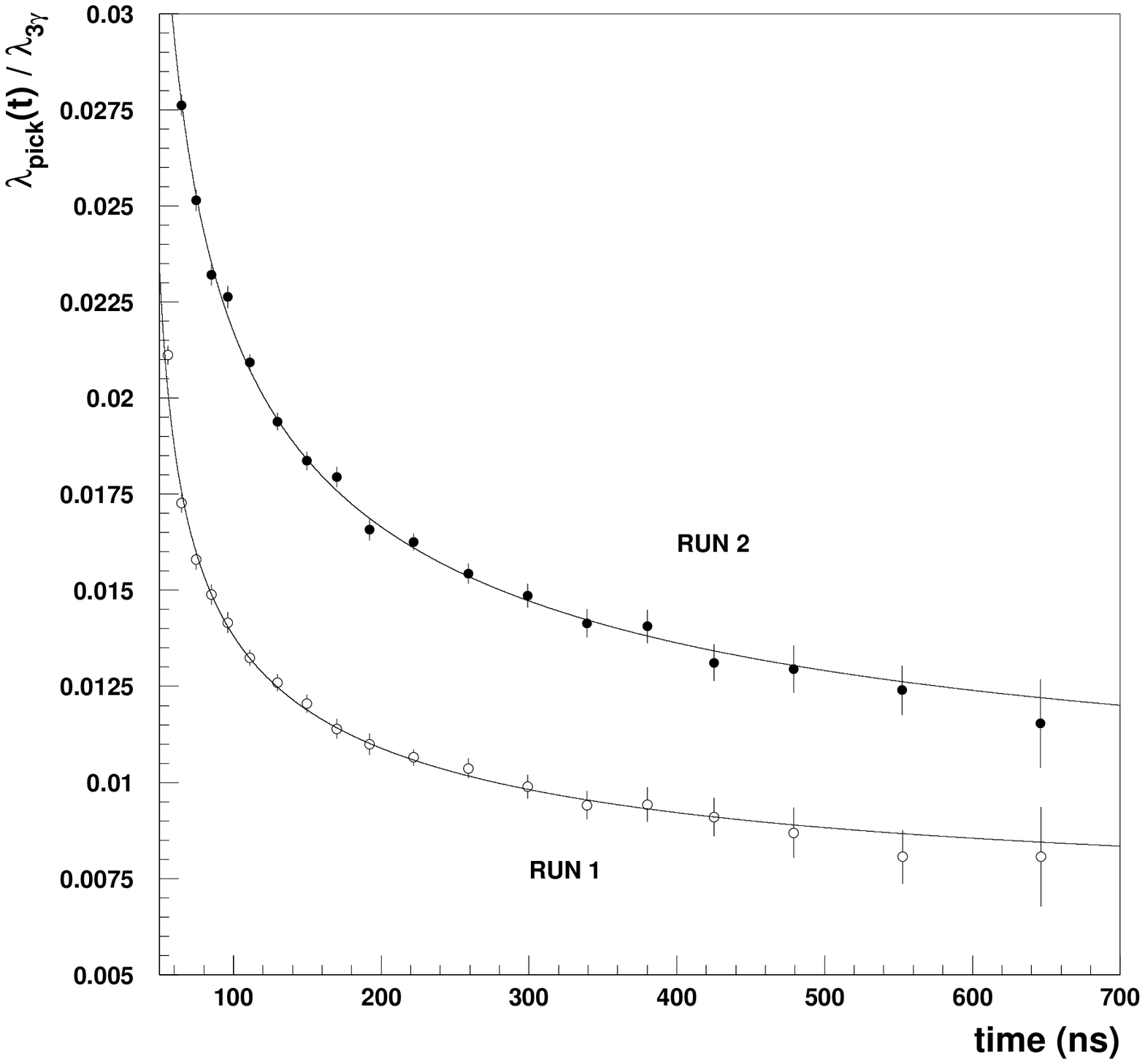}
  \end{center}
\caption{The ratio $\lambda_{pick}(t)/\lambda_{3\gamma}$ are plotted as a function of 
time. Open circles are data points for RUN1 and closed circles for RUN2. Solid lines 
represent best fit results obtained using Eq.~\ref{eq:therm2}. 
}
\label{fig:pick_ratio}
\end{figure}

These calculations of the $\lambda_{pick}/\lambda_{3\gamma}$ are
performed for various time windows,
and its time dependence is shown in Fig.~\ref{fig:pick_ratio}\@.
Since the fractional energy loss of o-Ps per collision with 
${\rm SiO_2}$ powder and the collision rate are both dependent 
on its energy, the time dependence of the 
average kinetic energy of o-Ps at time $t$, $\overline{E(t)}$ 
can be derived from the Boltzmann equation 

\begin{equation}
\frac{d}{dt}\overline{E(t)}=-\sqrt{2m_{Ps}\overline{E(t)}}\left(\overline{E(t)}-
\frac{3}{2}k_BT\right)\sum^\infty_{j=0}a_j\left(\frac{\overline{E(t)}}
{k_BT}\right)^{j/2},
\label{eq:therm1}
\end{equation}

where $m_{Ps}$ is the mass of o-Ps, $T$ room temperature, 
and $k_B$ the Boltzmann constant. 
The last term, the momentum transfer cross-section of ${\rm SiO_2}$ is
expanded in terms of $\overline{E(t)}$, i.e., 
the coefficients $a_j$ represent the effect of effective mass 
at the surface of the ${\rm SiO_2}$ grain and 
mean distance between the grains. Since the pick-off rate 
is proportional to the average velocity of the o-Ps, 
the ratio $\theta(t)\equiv\lambda_{pick}(t)/\lambda_{3\gamma}$
can be expressed by a differential 
equation\cite{SIO2}, i.e.,

\begin{equation}
\frac{d}{dt}\theta(t)=-C\left(\theta(t)^2-\theta_\infty^2\right)\theta(t)^{2\beta},
\label{eq:therm2}
\end{equation}

where $C$ is a constant, $\theta_\infty\equiv\theta(t\rightarrow\infty)$, and the last 
summation term in Eq.~\ref{eq:therm1} is replaced with an arbitrary real number 
$\beta$\@.
The measured $\lambda_{pick}(t)/\lambda_{3\gamma}$'s are fit with this equation
and figure~\ref{fig:pick_ratio} shows best fit results using the MINUIT library,
where the pick-off rate cannot be assumed as constant, even in ${\rm 
SiO_2}$ powder where the collision rate is expected to be higher. 

\begin{figure}[t!]%
\vspace{0.2pc}
  \begin{center}
    \includegraphics[width=20pc,keepaspectratio]{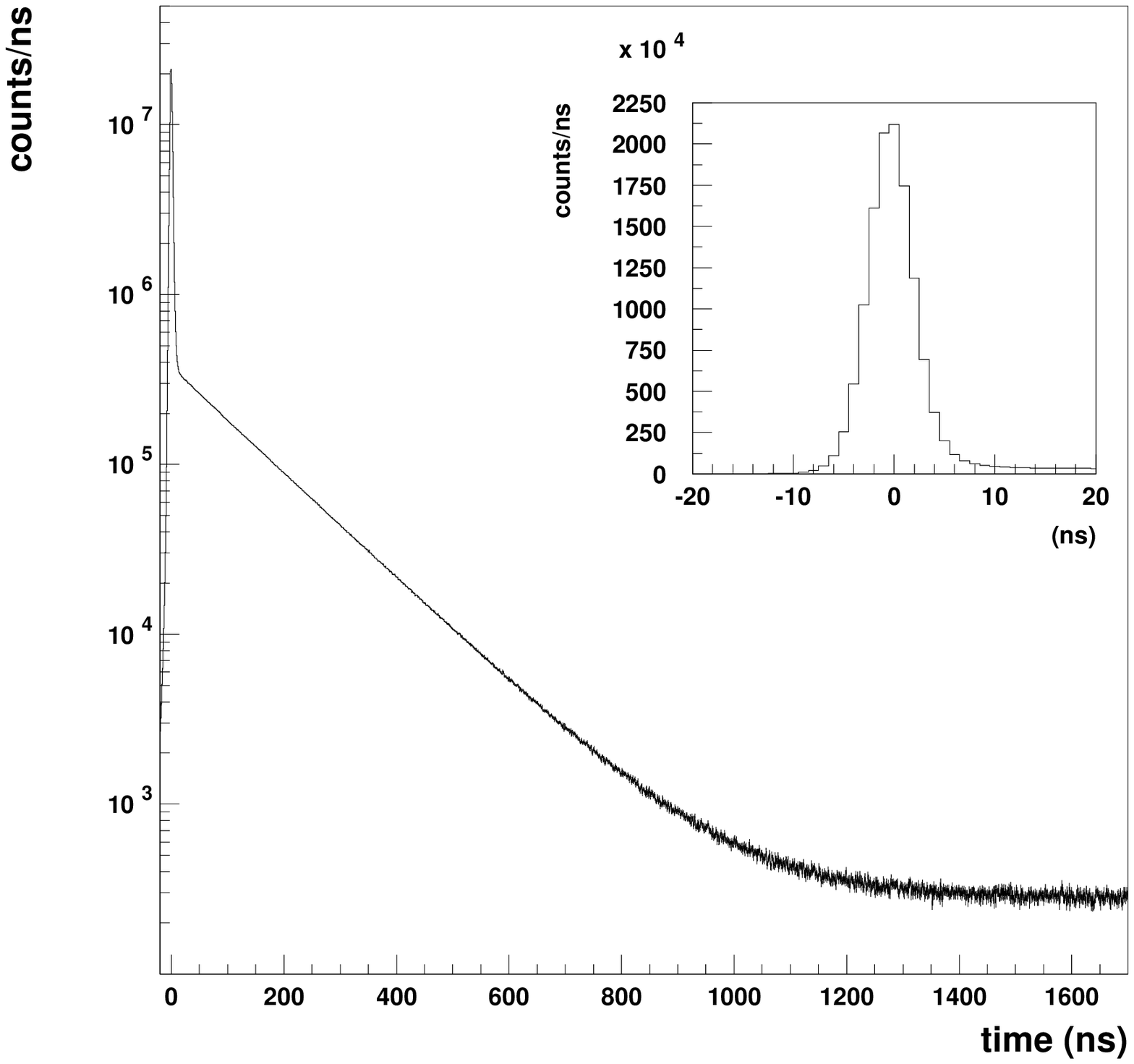}
  \end{center}
\caption{Time spectrum of NaI(Tl) scintillators for RUN 1 within an energy window of 
$370-440$~keV. The inset shows an enlarged view of the prompt peak, where good time 
resolution of $\sigma=2.2~ns$ is obtained.}
\label{fig:nai_timespectrum}
\end{figure}
Figure~\ref{fig:nai_timespectrum} shows the time spectrum of NaI(Tl) scintillators for 
RUN 1 with an energy window of $370-440$~keV, in which a sharp peak of the prompt 
annihilation is followed by the exponential decay curve of o-Ps and then the constant 
accidental spectrum.
The o-Ps curve is widely observed over about $1.2~\mu s$ corresponding
to about eight times the o-Ps lifetime, 
because of the weak positron source used in this experiment and 
good suppression of accidental contributions by selecting
the $\gamma$-energy. 
Observation in the wide range is important to understand time-dependence of $\lambda$\@.
To effectively eliminate pile-up events, a base-line 
cut condition was applied, and to further reject pile-up events, events with small 
differences between two ADC values (wide and narrow gates) were selected.

We fit resultant time 
spectrum using the least square method, i.e.,
\begin{equation}
N_{obs}(t)=\exp(-R_{stop}t)\left[\left(1+\frac{\epsilon_{pick}}
{\epsilon_{3\gamma}}\frac{\lambda_{pick}(t)}{\lambda_{3\gamma}}\right)N(t)
+C\right], 
\label{eq:tspec_obs}
\end{equation} 
where $\epsilon_{pick}$ and $\epsilon_{3\gamma}$ are 
the detection efficiencies for pick-off 
annihilations and $3\gamma$ decays, and $R_{stop}$ 
is an experimental random counting rate 
representing the fact that time interval measurement 
always accept the first $\gamma$ as a stop signal. 
$\lambda_{pick}/\lambda_{3\gamma}$ is about $1\%$ 
due to the low-density of the ${\rm SiO_2}$ powder, i.e., 
the ratio of error propagation to decay rate is suppressed by a factor of 100. 

\begin{figure}[t!]%
\vspace{0.2pc}
  \begin{center}
    \includegraphics[width=20pc,keepaspectratio]{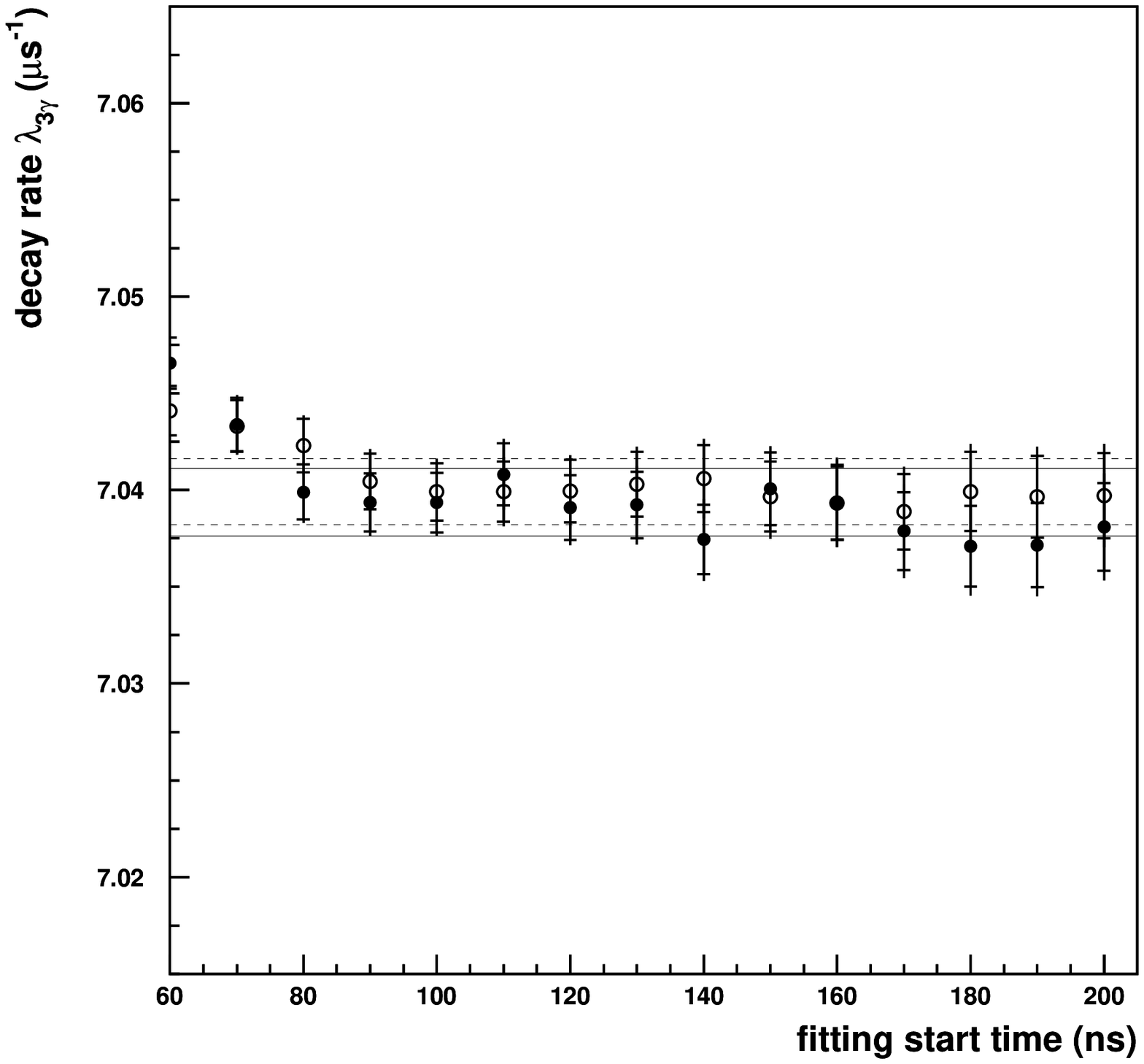}
  \end{center}
 \caption{Decay rates as a function of fitting start time. Small horizontal lines on the error 
bars represent the size of statistical errors solely due to fitting, while vertically extending 
bars include propagated errors from 
$\lambda_{pick}(t)/\lambda_{3\gamma}$ determination. Open and closed data points 
indicate values for RUN1 and RUN2, respectively. Dashed and solid lines show one 
standard deviation obtained at $t=100~ns$ for RUN1 and RUN2, respectively.}
\label{fig:fit_result}
\end{figure}

Figure~\ref{fig:fit_result} shows obtained fitting as a 
function of fitting start time for both runs, 
where values are stable with respect to fitting start time except those before $80~ns$. 
Since the fitting $\chi^2$'s for both runs rapidly increase 
before $80~ns$ due to the tail effect of prompt peak, values at 
$t=100~ns$ are taken as the final results. 
The reduced $\chi^2$'s at $t=100~ns$ are $1.006$ and 
$1.051$ for RUN 1 and RUN 2, respectively.
The fitted values of decay rate are extremely flat after 90 nsec, and
are confirmed to be independent of fitting start time\footnote{
The systematic increase in decay rate before $t_{start}=200~ns$ observed
in previous measurement\cite{ASAI95} was eliminated using the new system.}\@.
The obtained decay rates are 
$\lambda_{3\gamma}=7.03991\pm0.0017(stat.)~\mu s^{-1}$ for RUN 1 and 
$7.03935\pm0.0017(stat.)~\mu s^{-1}$ for RUN 2, 
which are consistent with each other. 
As shown in Fig.~3, the thermalization process and pickoff ratio are
different within each run,
the good correspondence between two runs is obtained. 
This indicates that our method correctly takes into account thermalization and 
pickoff correction. 

\renewcommand{\arraystretch}{1.00}
\begin{table}[htb]
\begin{center}
\begin{tabular}{lrr}\hline\hline
Source of Contributions&RUN1 ($ppm$)& RUN2 ($ppm$) 
\\\hline\hline
TDC module dependence & & \\
\quad-- Calibration & $<1$ & $<1$ \\
\quad-- Stability   & $2\sim3$ & $2\sim3$  \\
\quad-- Integral Non Linearity & $<~15$ & $<~15$ \\
\quad-- Differential Non Linearity & Negligible & Negligible \\\hline
Cut condition dependence & & \\  
\quad-- Base Line Selection &$-17$ and $+89$ & $-11$ and $+23$
 \\
\quad-- WD-NW condition & $-6$ and $+45$ & $-20$ \\ \hline
Monte Carlo dependence & & \\
\quad-- Normalization & $\pm99$ & $\pm113$ \\
\quad-- Relative efficiency  &  &  \\ 
\hspace*{15mm}of NaI(Tl) scintillator & $+7$ & $+7$\\
\quad-- Inhomogeneity of ${\rm SiO_2}$ powder & $<\pm55$ & $<\pm88$ \\\hline
Other Sources & & \\
\quad-- Zeeman effect & $-5$ & $-5$ \\
\quad-- Three-photon annihilation & $-26$ & $-42$ \\
\quad-- Stark effect & $-21$ & $-6$ \\
\quad-- $n=2$ excited state 	& $+19$ & $+19$\\\hline
Total  & $-120$ and $+153$ &  $-151$ and $+148$ \\ \hline\hline
\end{tabular}
 \caption{Summary of systematic errors. See text for details of each item.}
\label{table:systematic}
\end{center}
\end{table}

Table~\ref{table:systematic} summarizes estimates of various systematic errors, and 
the details are shown in reference\cite{D-thesis}.
We just mention here about errors related to simulation and Stark effect.

Monte Carlo simulation related errors:
the predominant contribution to total systematic error is produced 
by uncertain normalization. That is, the 
number of pick-off events are determined by subtracting 
the normalized 3-$\gamma$ spectrum of Monte 
Carlo simulation from the o-Ps spectrum, 
where changing the normalization factor affects the 
$\lambda_{pick}(t)/\lambda_{3\gamma}$ values and 
eventually propagates to the final result. 
Since the sharp fall-off of the 3$\gamma$-spectrum at 511~keV is solely 
produced by the good Ge energy resolution of $\sigma= 0.5$~keV, 
this subtraction only affects the lower side of the pick-off spectrum such that 
improper subtraction results in asymmetry of the pick-off spectrum shape. 
Comparison of the asymmetries of the pick-off peak shape 
and the prompt peak annihilation spectrum is a good parameter for estimating 
this systematic error. The 1~$\sigma$ error is assessed as $\pm99~ppm$\@.
Inhomogeneity of the powder affects on the detection efficiency.
${\rm SiO_2}$ powder density in the MC simulation is conservatively 
changed by $\pm10\%$ although the uniformity is known to be within a few \%; 
a change resulting in an error of $<\pm55$~ppm. 

The Stark shift stretches the lifetime of Ps atoms, i.e., 
a perturbative calculation shows that the shift is 
proportional to a square of the effective electric field $E$ such that 
$\triangle\lambda_{3\gamma}/\lambda_{3\gamma}=248\cdot(E/E_0)^2$, 
where $E_0 = m_e^2e^5/\hbar^4 \approx 5.14\times10^9~V/cm$. 
$E$ is defined as the root-mean-square electric field sensed by o-Ps during its lifetime. 
Calculations have estimated two contributions exist based on 
measurements of the electrical charge-up on the primary grains of 
silica powders and electrical dipole moment on the surface of grains\cite{D-thesis}. 
The charge-up is partly intrinsic depending on powder 
specifications and partly due to positron depositions from the $\beta^+$ source. 
The effect, however, is negligible in both cases, 
i.e., on the level of $10^{-2}~ppm$ at most. 
Silanol functional groups on the surface of the powder grain 
behave as an electrical dipole moment creating an effective field around the grains. 
Average densities are measured to be  $2.5 /nm^2$ and $0.44 /nm^2$ for RUN 1 and 2, 
respectively. 
Accordingly, the effective field can be analytically calculated 
such that the contribution to the o-Ps decay 
rate is determined to be $-21~ppm$ for RUN1 and $-6~ppm$ for RUN2. 
These estimations were confirmed  using results from precise hyper-fine-structure (HFS)
interval measurements of ground state Ps in silica powder\cite{HFS}, 
where the interval is proportional to the size of Stark effect. 
Considering the difference in powder densities used, 
the HFS results are consistent with our estimation.

Other sources of systematic errors:
Error contribution due to the Zeeman effect is estimated
using the measured absolute magnetic field around the positronium assembly ($-5~ppm$). 
Since the 3-$\gamma$ pick-off process can only occur at a certain 
ratio, the calculated relative frequency 
$\sigma_{3\gamma}/\sigma_{2\gamma}\sim1/378$ is consistent 
with previous measurements\cite{3-pick}, 
being $-26~ppm$ for RUN1 and $-42~ppm$ for RUN2. 
The  probability of the excited state (n=2) of Ps 
is about $3\times10^{-4}$ \cite{EXCITED}, which could make 
the intrinsic decay rate $19~ppm$ smaller 
due to the low decay rate ($\frac{1}{8}\lambda_{3\gamma}$) 

The above discussed systematic errors are regarded 
as independent contributions such that the total 
systematic error can be calculated as their quadratic sum, 
resulting in $-120~ppm$, $+153~ppm$ for RUN1 
and $-151~ppm$, $+148~ppm$ for RUN2.

\section{Conclusions}\label{sec:conclusion}
\begin{figure}[t!]%
  \begin{center}
    \includegraphics[width=20pc,keepaspectratio]{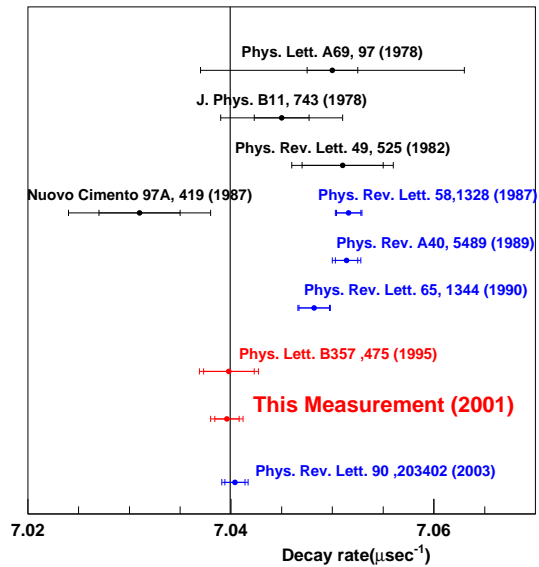}
  \end{center}
\vspace{-1.0pc}
 \caption{Historical plot of o-Ps decay rate measurements including present results. 
Vertical line shows the $O(\alpha^2)$-corrected NRQED prediction. 
Small vertical lines on the error bars indicate the size of errors solely due to statistics 
while error bars represent total ambiguities including systematic errors.}
\label{fig:history}
\end{figure}

The decay rate of o-Ps formed in ${\rm SiO_2}$ was measured using
a direct $2\gamma$ correction method in which the thermalization 
effect of o-Ps is accounted for and integrated into the time spectrum 
fitting procedure. 
Consistent results were obtained using two runs with different types of ${\rm SiO_2}$
powders, i.e., and a weighted average gave $\lambda_{\rm o-
Ps}=7.0396\pm0.0012(stat.)\pm0.0011(sys.)\mu s^{-1}$, 
which agrees well with our previous result\cite{ASAI95,JIN00}, 
it disagrees with results obtained at Ann Arbor, 
i.e., by 3.8--$5.6~\sigma$ \cite{GAS89,CAV90}\@. 
As illustrated in Fig.~\ref{fig:history}, our value agrees well with the 
NRQED prediction corrected up to $O(\alpha^2)$ term\cite{ADKINS-4}. 
We confirm our results obtained at 1995 and 2000,
and improve accuracy by factor of 1.8\@.

Just one week before this workshop,
Ann Arbor reported new result\cite{baka}
using nanoporous silica film, which produces
near-thermal energy o-Ps. 
They recognized the systematic problem due to unthermalized o-Ps (fast o-Ps) and 
suppress the kinetic energy of the emitted o-Ps using nanoporous silica film and positron 
beam energy.
They obtain value of $\lambda_{\rm o-Ps}=7.0404\pm0.0010(stat.)\pm0.0008(sys.)\mu s^{-1}$,
which completely agrees with our results as shown in Fig.6.
So we can conclude that 
the unthermalized o-Ps has contributed large systematic errors
of about 1000 ppm and made discrepancy between the QED prediction
as pointed out in references\cite{PEKIN,ASAI95}.
Now two groups (two methods) obtain the consistent results
and the orthopositronium lifetime puzzle is clearly solved here.
The combined results of two different methods is $7.0400(10) \mu s^{-1}$,
which is not enough accuracy(140ppm) to examine the $O(\alpha^2)$-correction (210ppm).
We need much effort to improve accuracy.

We wish to thank Prof. S. Orito, T. Hyodo and Y. Nagashima (Univ. of Tokyo) for 
very useful suggestions for discussions.
It is a pleasure to thank Prof. G.S. Adkins (Franklin \& Marshall
College) for calculating the $O(\alpha)$ corrected energy spectrum.


\begin{thebibliography}{00}

\bibitem{GAS87}
{C. I. Westbrook, D. W. Gidley, R. S. Conti, and A. Rich}, {\it Phys. Rev. Lett}. {\bf 58}
  1328 (1987).

\bibitem{GAS89}
{C. I. Westbrook, D. W. Gidley, R. S. Conti, and A. Rich}, {\it Phys. Rev.} {\bf A40}
 5489 (1989).

\bibitem{CAV90}
{J. S. Nico, D. W. Gidley, A. Rich, and P. W. Zitzewitz}, {\it Phys. Rev. Lett.} {\bf 65}
1344 (1990).

\bibitem{ADKINS-4}
{G. S. Adkins, R. N. Fell, and J. Sapirstein}, {\it Phys. Rev. Lett.} {\bf 84} 5086 (2000) and
{\it Ann. Phys.} {\bf 295} 136 (2002).

\bibitem{EXOTIC-LL}
{S. Asai, S. Orito, K. Yoshimura, and T. Haga}, {\it Phys. Rev. Lett.} {\bf 66}  2440 (1991); \\
{S. Orito, K. Yoshimura, T. Haga, M. Minowa, and M. Tsuchiaki},
{\it Phys. Rev. Lett.} {\bf 63} 597 (1989).

\bibitem{EXOTIC-SL}
{T. Maeno, M. Fujikawa, J. Kataoka, Y. Nishihara, S. Orito, K. Shigekuni, Y.
  Watanabe}, {\it Phys. Lett.} {\bf B351} 574 (1995); \\
{S. Asai, K. Shigekuni, T. Sanuki, and S. Orito}, {\it Phys. Lett.} {\bf B323} 90 (1994); \\
{M. Tsuchiaki, S. Orito, T. Yoshida, and M. Minowa}, {\it Phys. Lett.} {\bf B236} 81 (1990)
 81.

\bibitem{EXOTIC-IV}
{T. Mitsui, R. Fujimoto, Y. Ishisaki, Y. Ueda, Y. Yamazaki, S. Asai, and S.
  Orito}, {\it Phys. Rev. Lett.} {\bf 70} 2265 (1993).

\bibitem{EXOTIC-UB}
{T. Mitsui, K. Maki, S. Asai, Y. Ishisaki, R. Fujimoto, N. Muramoto, T. Sato,
  Y. Ueda, Y. Yamazaki and S. Orito}, {it Euro. phys. Lett.} {\bf 33} 111 (1996); \\
{A. Badertscher, P. Crivelli, M. Felcini, S.N. Gninenko, N.A. Goloubev, P. Nedelec, 
J.P. Peigneux, V.Postoev, A. Rubbia and D. Sillou} {\it  Phys. Lett. } {\bf B542} 
29 (2002) 

\bibitem{EXOTIC-TW}
{S. Asai, S. Orito, T. Sanuki, M. Yasuda, and T. Yokoi}, {\it Phys. Rev. Lett.} {\bf 66}
1298 (1991); \\
{D. W. Gidley, J. S. Nico, and M. Skalsey}, {\it Phys. Rev. Lett.} {\bf 66} 1302 (1991).

\bibitem{EXOTIC-FOUR}
{K. Marko and A. Rich}, {\it Phys. Rev. Lett.} {\bf 33} 980 (1974).

\bibitem{PEKIN}
{S. Asai , T. Hyodo, Y. Nagashima, T.B. Chang and S. Orito}, {\it Materials 
Science Forum,} {\bf 619} 175 (1995).

\bibitem{ASAI95}
{S.Asai} {`` New measurement of orthopositronium lifetime''}, Ph. D. 
thesis, University of Tokyo (1994);\\
{S. Asai, S. Orito, and N. Shinohara}, {\it Phys. Lett.} {\bf B357} 475 (1995).

\bibitem{JIN00}
O. Jinnouchi, S. Asai, T. Kobayashi,  hep-ex/0011011 (2000).

\bibitem{D-thesis}
{O.~Jinnouchi}, ``Study of bound state QED: precision measurement of the
  orthopositronium decay rate'', Ph. D. thesis, University of Tokyo (2001);\\
O. Jinnouchi, S. Asai, T. Kobayashi ``Precision Measurement of Orthopositronium 
Decay Rate Using ${\rm SiO_2}$ Powder '' accepted by  {\it Phys. Lett. B}\@.  

\bibitem{Ann1}
M.Skalsey et al,  {\it Phys. Rev. Lett.} {\bf 80} 3727 (1998).

\bibitem{SIO2}
{Y. Nagashima, T. Hyodo, K. Fujiwara, and A. Ichimura}, {\it J. Phys.} {\bf B31} 329
(1998);\\
{Y. Nagashima, M. Kakimoto, T. Hyodo, K. Fujiwara, A. Ichimura, T. Chang, J.
  Deng, T. Akahane, T. Chiba, K. Suzuki, B. T. A. McKee, and A. T. Stewart},
{\it Phys. Rev. }{\bf A 52} 258 (1995).

\bibitem{3-pick}
{J.~A.~Rich}, {\it Phys.~Rev} {\bf 61} 140 (1951).

\bibitem{HFS}
{M.H.~Yam, P.O.~Egan, W.E.~Frieze, and V.M.~Hughes}, {\it Phys.~Rev.} {\bf A18} 350 (1978). 

\bibitem{EXCITED}
{S. Hatamian, R. S. Conti, and A. Rich}, {\it Phys.~Rev.~Lett.} {\bf 58} 1833 (1987).

\bibitem{Adkins_Private}
{G.S.~Adkins}, private communication.

\bibitem{baka}
{R. S. Vallery, P. W. Zitzerwitz and D. W. Gidley}, {\it Phys.~Rev.~Lett.} {\bf 90} 203402 (2003).

\end{thebibliography}
\end{document}